\documentclass[prl,reprint,floatfix,superscriptaddress]{revtex4-1}
\usepackage{times}
\usepackage{graphicx}
\usepackage{hyperref}	
\usepackage{amsmath,amssymb}
\usepackage{physics}
\usepackage{nicefrac} 
\usepackage{color}

\newcommand{\SiVm}[0]{SiV$^-$}
\newcommand{\NVm}[0]{NV$^-$}

\newcommand{\StandardSample}[0]{Sample 13} 
\newcommand{\PurifiedSample}[0]{Sample 12}

\begin{document}

\title{
Silicon-Vacancy Spin Qubit in Diamond: \\A Quantum Memory Exceeding 10 ms with
Single-Shot State Readout
}

\author{D. D. Sukachev}

\email[These authors contributed equally to this work\\]{sukachev@fas.harvard.edu}
\affiliation{Department of Physics, Harvard University, 17 Oxford Street, Cambridge, Massachusetts 02138, USA}

\author{A. Sipahigil}
\email[These authors contributed equally to this work\\]{sukachev@fas.harvard.edu}

\affiliation{Department of Physics, Harvard University, 17 Oxford Street, Cambridge, Massachusetts 02138, USA}

\author{C. T. Nguyen}
\email[These authors contributed equally to this work\\]{sukachev@fas.harvard.edu}
\affiliation{Department of Physics, Harvard University, 17 Oxford Street, Cambridge, Massachusetts 02138, USA}

\author{M. K. Bhaskar}
\affiliation{Department of Physics, Harvard University, 17 Oxford Street, Cambridge, Massachusetts 02138, USA}

\author{R. E. Evans}
\affiliation{Department of Physics, Harvard University, 17 Oxford Street, Cambridge, Massachusetts 02138, USA}

\author{F. Jelezko}
\affiliation{Institute for Quantum Optics, Ulm University and Center for Integrated  Quantum Science and Technology, Albert-Einstein-Allee 11, 89081 Ulm, Germany}

\author{M. D. Lukin}
\email{lukin@physics.harvard.edu}
\affiliation{Department of Physics, Harvard University, 17 Oxford Street, Cambridge, Massachusetts 02138, USA}

\begin{abstract}
The negatively-charged silicon-vacancy (\SiVm) color center in diamond 
has recently emerged as a promising system for quantum photonics.  
Its symmetry-protected optical transitions enable creation of indistinguishable emitter arrays and deterministic
coupling to  nanophotonic devices. 	
Despite this, the longest coherence time associated with its electronic spin achieved to date ($\sim 250$\,ns) has been limited  by coupling to acoustic phonons.
We demonstrate coherent control and suppression of  phonon-induced dephasing of the \SiVm{} electronic spin coherence by five orders of magnitude by operating at temperatures below 500\,mK.
By aligning the magnetic field along the \SiVm{} symmetry axis, we demonstrate spin-conserving optical transitions and single-shot readout of the \SiVm spin with 89\% fidelity.
Coherent control of the \SiVm spin with microwave fields  is used to  demonstrate a spin coherence time $T_2$ of 13\,ms and a spin relaxation time $T_1$ exceeding 1\,s  at 100\,mK. 
These results establish the \SiVm as a promising solid-state candidate for the realization of quantum networks. 
\end{abstract}

\maketitle

Quantum networks require the ability to store quantum information in long-lived memories, to efficiently interface these memories with optical photons and to provide quantum nonlinearities required for deterministic  quantum gate operations ~\cite{briegel1998quantum, Childress2006a}.
Even though key building blocks of quantum networks have been  demonstrated in various 
systems~\cite{ritter2012elementary,hucul2015modular}, no solid-state platform has satisfied these requirements.
Over the past decade, solid-state quantum emitters with stable spin degrees of freedom such as charged quantum dots and nitrogen-vacancy (NV) centers in diamond have been investigated for the realization of quantum network nodes~\cite{Gao2015}. 
While quantum dots can be deterministically interfaced with optical photons~\cite{Lodahl2015},
 their quantum memory time is limited to the $\mu$s scale ~\cite{Press2010} 
due to interactions with their surrounding nuclear spin bath. 
In contrast, NV centers have  an exceptionally  long-lived  quantum memory~\cite{Balasubramanian2009} but suffer from weak, spectrally unstable optical transitions~\cite{Faraon2012}. 
Despite impressive proof-of-concept experimental demonstrations with these systems~\cite{hensen2015loophole,delteil2015generation}, scaling to a large number of nodes is limited by the  challenge of identifying  suitable quantum  emitters with the combination of strong, homogeneous and coherent optical transitions and long-lived quantum memories.

The negatively-charged silicon-vacancy (\SiVm ) has recently been shown to have bright, narrowband optical transitions with a small inhomogeneous broadening~\cite{neu2011single,Rogers2014}. 
The optical coherence of the \SiVm{} is protected by its inversion symmetry~\cite{Sipahigil2014}, even  in nanostructures~\cite{Evans2016}. 
These optical properties were recently used to show strong interactions between single photons and single \SiVm{} centers and to probabilistically entangle two \SiVm{} centers in a single nanophotonic device~\cite{Sipahigil2016}.
At 4\,K, however, the \SiVm{} spin coherence  is limited to $\sim 100$\,ns due to coupling to the phonon bath, mediated by the spin-orbit interaction~\cite{Rogers2014a,Pingault2014,Jahnke2015,Becker2016a,Pingault2017}.

\begin{figure}[b!]
	\includegraphics[width=\linewidth]{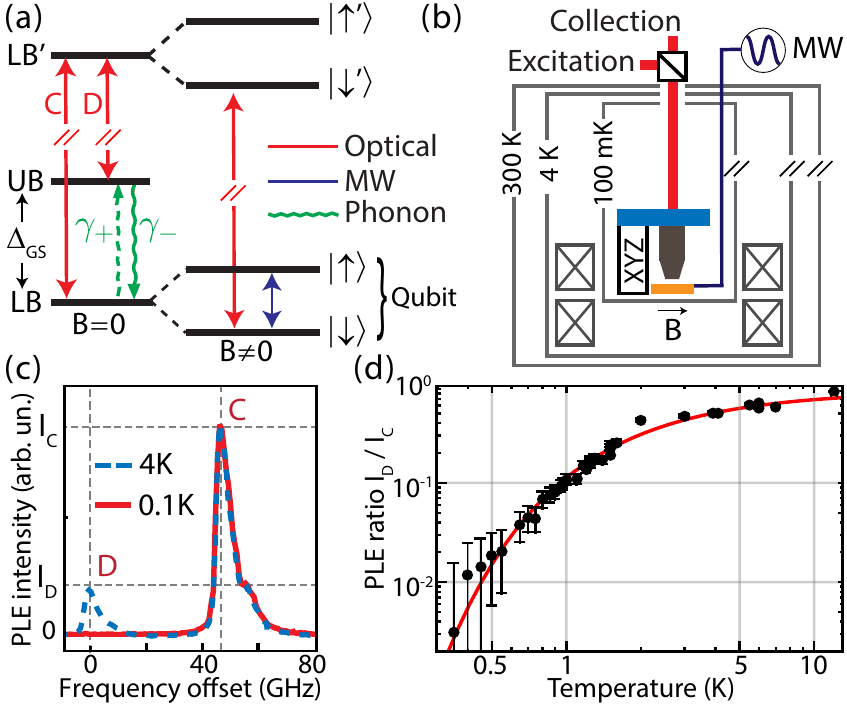}
		\caption{\label{fig1}
		(a) \SiVm\ electronic structure. Optical transitions C and D connect the lower (LB) and upper (UB) spin-orbit branches to the lowest-energy optical excited state $\left(\text{LB}'\right)$. 
	Each branch is split into two spin sublevels in a magnetic field $\vec{B}$. Red and blue arrows denote optical and microwave transitions, respectively. $\gamma_+$ and $\gamma_-$ are phonon-induced decay rates.  
	 (b) Schematic of the setup. An objective is mounted on piezo positioners to image the diamond sample using free-space optics.  The combined system is attached to the mixing plate of a dilution refrigerator and placed inside a superconducting vector magnet. 
	(c) PLE spectra of an \SiVm{} ensemble at $B = 0$ for $T=4$\,K and $0.1$\,K. The peak intensity $I_C$ ($I_D$) is proportional to the population in the LB (UB). (d) 
	$I_D / I_C$ (and $\gamma_+/\gamma_-$) is reduced at low temperatures, following $e^{-h\Delta / k_B T}$ with $\Delta_{\textrm{fit}}=42 \pm 2$\,GHz in agreement with the measured $\Delta_{\textrm{GS}}=48$\,GHz.
	}
\end{figure}

In this Letter, we demonstrate high-fidelity coherent manipulation and  single-shot readout of individual  \SiVm spin qubits in a dilution refrigerator. In particular, we extend the coherence time of the \SiVm{} electronic spin by five orders of magnitude to $13$\,ms by operating at $100$\,mK~\footnote
{
The neutral SiV center (SiV$^0$) features suppressed phonon-induced dephasing of its spin state~\cite{Edmonds2008} and [Green, {\it et. al.},  arXiv:1705.10205]
yielding coherence times of $T_2 \sim 200$\,ms in bulk measurements at $T = 15$\,K~[Rose, {\it et. al.}, arXiv:1706.01555].
However, their optical properties, including radiative quantum efficiency, are currently believed to be inferior to those of \SiVm~[U. F. S. D’Haenens-Johansson, {\it et. al.}, Physical Review B {\bf 84}, 245208 (2011)], making it challenging to   realize  an efficient photonic interface.
}.

The key idea of the present work can be understood by considering the energy level diagram of the \SiVm~[Fig.~1(a)]. 
The ground state of the \SiVm{} is split by the spin-orbit interaction and crystal strain into a lower branch (LB) and an upper branch (UB)
 separated by $\Delta_\text{GS}$.
Each branch comprises two degenerate spin sublevels~\cite{hepp2014}.
Application of a magnetic field lifts the spin degeneracy and allows the use of the spin sublevels $\ket{\downarrow}$ and $\ket{\uparrow}$ of the LB as qubit states.
At 4\,K, the \SiVm{} spin coherence is limited to $\sim 100$\,ns ~\cite{Pingault2014,Rogers2014a,Jahnke2015,Becker2016a,Pingault2017} due to interactions with the thermal acoustic phonon bath at frequency $\Delta_\text{GS}\sim 50$\,GHz. 
These interactions result in a relaxation at rates $\gamma_+$ and $\gamma_-$ between the levels in the LB and the UB with different orbitals and the same spin projections as shown in Fig.~1(a) and destroy spin coherence.
By reducing the occupation of phonon modes at $\Delta_{\textrm{GS}}$ at lower temperatures, one can  suppress the rate $\gamma_+$, leaving the spin qubit in a manifold free from phonon-induced decoherence, thereby increasing spin coherence ~\cite{Jahnke2015} .

We investigate the \SiVm\ properties below $500$\,mK using a dilution refrigerator with a free-space confocal microscope and a vector magnet as shown in Fig.~1(b). 
Details of the experimental setup are available in the Supplemental Material~\cite{SI}.
We first study the thermal population of the LB and the UB between $0.1$ and $10$\,K using an ensemble of as-grown \SiVm centers (Sample-A in Ref.~\cite{Edmonds2008}). 
We probe the relative populations in the LB and the UB by measuring the ensemble photoluminescence excitation (PLE) spectra of transitions C and D.
Transitions C and D are both visible in PLE at 4\,K, which suggests comparable thermal population in the LB and UB [Fig.~1(c)]. 
As the temperature is lowered [Fig.1(d)], the ratio of the transition D and C peak amplitudes ($I_D / I_C$) reduces by more than two orders of magnitude and follows $e^{-h \Delta_\text{GS} / k_B T}$~\cite{Jahnke2015}. These measurements demonstrate an orbital polarization in the LB of $> 99\%$ below $500$\,mK. At these low temperatures, $\gamma_+ << \gamma_-$ and the qubit states are effectively decoupled from the phonon bath.

To investigate the coherence properties of single emitters, we create single \SiVm{} centers at a depth of $\sim250$\,nm via $^{28}$Si ion implantation at a dose of $10^9$\,cm$^{-2}$ and an energy of 380\,keV into two type-IIa ([N]\,$< 5$\,ppb, [B]\,$<1$\,ppb) diamond samples (Element Six).
The first  sample (\StandardSample) has a natural abundance of 1.1\% of $^{13}$C isotopes with a nuclear spin $I=1/2$. The second sample (\PurifiedSample) is engineered to have only $10^{-3}\%$ $^{13}$C to suppress  hyperfine interactions between the spin qubit and  the nuclear spin bath \cite{Balasubramanian2009}. 
After ion implantation and high-temperature annealing~\cite{Evans2016}, we fabricate a shorted coplanar waveguide on the diamond to drive microwave (MW) transitions between the qubit states~\cite{SI}.

We use spin-selective optical transitions between states $\ket{i}$ and $\ket{i'}$ at frequencies f$_{ii'}$ $\left(i=\{\uparrow,\downarrow\}\right)$ [Fig.~2(a)] to optically initialize and readout the qubit states. 
Applying a magnetic field $B \sim 0.5-3$\,kG  allows us to optically resolve these transitions.
Fig.~2(b) shows the PLE spectrum of the spin-selective optical transitions at $4$\,K (red circles). 
These resonances disappear in continuous wave measurements at 100\,mK (blue squares). 
This effect results from optical pumping of the  qubit to the long-lived dark spin state. 
The central peak originates from off-resonant scattering from the two spin transitions.

\begin{figure}
	\includegraphics{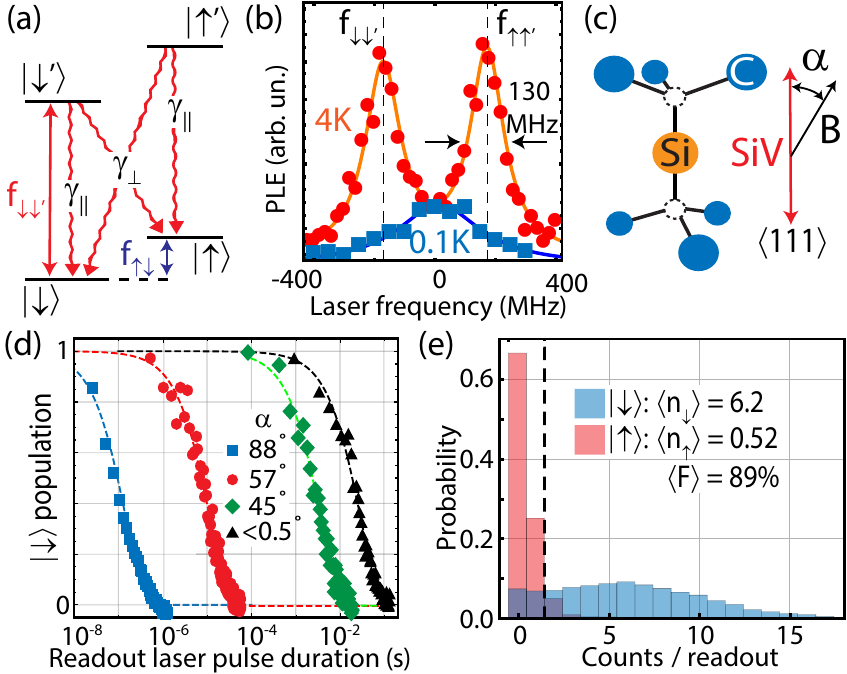}
		\caption{\label{fig2}
	(a) Spin-selective optical transitions and branching ratios. f$_{ij}$ is the transition frequency between states $i$ and $j$. 
	$\gamma_{||}$ and $\gamma_{\perp}$ are spin-conserving and spin-flipping decay rates, respectively; $f_{\uparrow \downarrow}$ is the qubit frequency.
	(b) PLE spectra measured 	at 4\,K and 0.1\,K. 
	(c) Schematic of the \SiVm\ molecular structure \cite{hepp2014}. 
	$\alpha$ is the angle between the magnetic field $\vec{B}$ and the SiV symmetry axis set by the two lattice vacancies (empty circles) and aligned along the $\left\langle 111 \right\rangle$ diamond axis.
	(d) Optical spin pumping timescale $\tau$ measured at different $\vec{B}$. 
	Here, $\{\alpha, B, \tau\}=$
	$\{88^\circ, 2.9$\,kG$, 140$\,ns$\}$ for the blue squares; 
	$\{57^\circ, 3.0$\,kG$, 10$\,$\mu$s$\}$ for the red circles; 
	$\{45^\circ, 1.7$\,kG$, 3$\,ms$\}$ for the green diamonds and  
	$\{ <0.5^\circ, 2.7$\,kG$, 30$\,ms$\}$ for the black triangles.
	(e) 
	Single-shot spin readout in $B=2.7$\,kG. A 20-ms long laser pulse at frequency f$_{\downarrow\downarrow'}$ is used for state readout. A second laser initializes the spin states via optical pumping. Spin readout photon statistics after initialization in state $\ket{\uparrow}$ (red) and $\ket{\downarrow}$ (blue). Average fidelity $F=89\%$.    
}
\end{figure}

To achieve high-fidelity readout of the spin states, it is desirable to scatter photons many times without causing a spin-flip~\cite{Robledo2011,Delteil2014}. To obtain such spin-conserving optical transitions, the cyclicity of the transition $\gamma_\parallel/(\gamma_\perp+\gamma_\parallel)$ can be tuned by varying the angle $\alpha$ of the applied magnetic field with the SiV symmetry axis as shown in Fig.~2(c)~\cite{Rogers2014a}. 
Figure~2(d) shows the optical spin pumping timescale for different $\alpha$ when the transition f$_{\downarrow\downarrow'}$ is driven near saturation. We extend the optical pumping timescale by more than five orders of magnitude from $100$\,ns for $\alpha \sim 90^{\circ}$ to $30$\,ms in an aligned field. 

The ability to optically excite the \SiVm\ $\sim 10^5$ times without causing a spin-flip ~\cite{SI} enables high-fidelity single-shot readout of the spin state despite the low photon collection efficiency $\left(\sim 10^{-4}\right)$ in the phonon-sideband (PSB).
We measure the spin state by driving the f$_{\downarrow\downarrow'}$ transition near saturation and monitoring fluorescence on the PSB. 
Fig.~2(e) shows the readout counts distributions for the spin initialized in state $\ket{\downarrow}$ (blue histogram) and $\ket{\uparrow}$ (red histogram) using a $150$\,ms-long pulse from a second laser at frequency f$_{\uparrow\uparrow'}$ or f$_{\downarrow\downarrow'}$, respectively. 
We detect $\langle n_{\downarrow} \rangle=6.2$ photons from state $\ket{\downarrow}$ and
$\langle n_{\uparrow} \rangle=0.52$ from state $\ket{\uparrow}$ in a 20-ms-long readout window. 
By choosing a state-detection threshold of $n>1$ for state $\ket{\downarrow}$ and $n\leq1$ for state $\ket{\uparrow}$, we obtain an average readout fidelity of $F=(F_\uparrow+F_\downarrow)/2=0.89$ where $F_i$ is the readout fidelity for state $i$. 
For the measurements in Figs.~3 and 4, we roughly align the magnetic field with $\alpha<5^{\circ}$ to operate in an efficient spin readout regime but do not optimize for the highest fidelity at each point.   
Under these conditions, we measure lifetimes ($T_1$) of the  qubit states exceeding 1\,s at 100\,mK~\cite{SI}.

The spin readout time ($\sim 10$\,ms) is currently  limited by the low collection efficiency of the setup~\cite{SI} and by optical pumping to the metastable UB of the ground state~\cite{SI} with a lifetime of $\sim 200$\,ns~\cite{Jahnke2015}. This readout time can be reduced by several orders of magnitude by adding a repumping laser on transition D [Fig.~1(a)] and by using nanophotonic structures to improve the collection efficiency to above 10\%~\cite{Sipahigil2016}.

\begin{figure}[t!]
	\includegraphics[width=\linewidth]{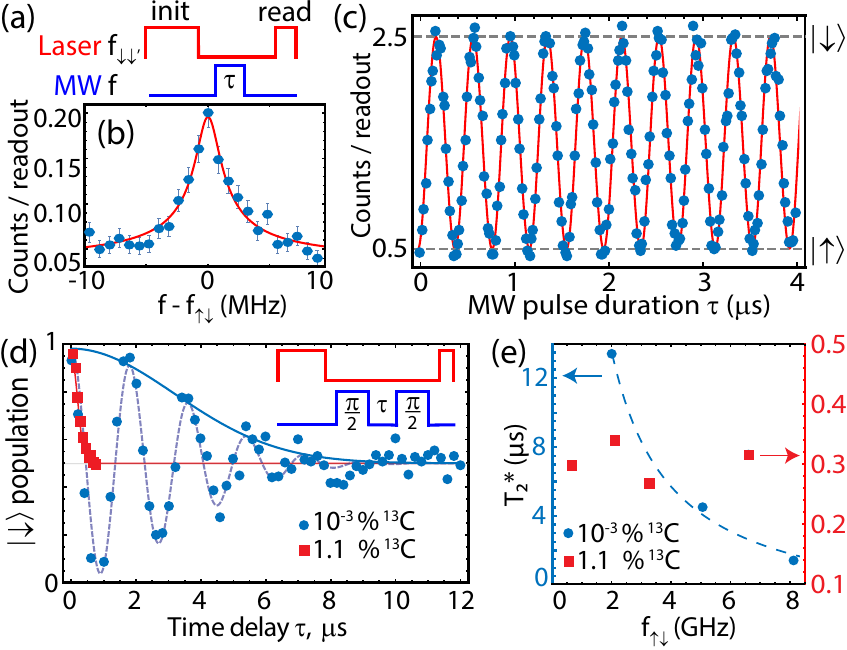}
		\caption{\label{fig3}
	(a) Pulse sequence for ODMR and Rabi measurements. 
	(b) Pulsed ODMR measurement for $\tau=500\,\mu$s. 
	Durations of the initialization and readout laser pulses are 15\,ms and 2\,ms, respectively~\cite{SI}.
	(c) Resonant driving at frequency f$_{\uparrow\downarrow}$ results in Rabi oscillations between states $\ket{\uparrow}$ and $\ket{\downarrow}$. 
	Data in (b) and (c) are from \PurifiedSample.
	(d) Ramsey interference measurement of $T_2^*$ for the two samples. MW pulses are detuned by $\sim 550$\,kHz from the $f_{\uparrow \downarrow }$ 
	for the blue 	data.  
	Duration of the initialization (readout) laser pulse is 15\,ms (2\,ms) for \PurifiedSample{} and 1.5\,ms (0.2\,ms) for \StandardSample.
	(e) $T_2^*$ as a function of qubit resonant frequency. Dashed blue line is a fit to $1/f_{\downarrow \uparrow}$ scaling~\cite{SI}.
	}
\end{figure}

To coherently control the \SiVm\ electron spin qubit we use a MW field at frequency f$_{\uparrow\downarrow}$~\cite{Pingault2017}. 
In the following experiments,  single strained \SiVm{} centers with $\Delta_\text{GS} \sim 80$\,GHz are used. 
When crystal strain is comparable to spin-orbit coupling ($\sim 48$\,GHz), the orbital components of the qubit states are 
no longer orthogonal~\cite{hepp2014}, leading to an allowed magnetic dipole transition~\cite{Pingault2017}.
This MW transition is allowed for both aligned and misaligned magnetic fields, 
allowing simultaneous MW control and single-shot readout of the \SiVm{} spin.

We focus on  single \SiVm{} centers placed less than $2\,\mu$m from the coplanar waveguide to efficiently drive the qubit transition with low MW powers and maintain a steady-state sample temperature below $100$\,mK. The spin qubit frequency $f_{\downarrow \uparrow}$ is determined using a pulsed optically-detected magnetic resonance (ODMR) measurement as shown in Figs.~3(a) and 3(b). 
A long laser pulse at frequency f$_{\downarrow\downarrow'}$ initializes the spin in state $\ket{\uparrow}$ via optical pumping. 
After a microwave pulse of duration $\tau$, a second laser pulse at 
f$_{\downarrow\downarrow'}$ 
reads out the population in state $\ket{\downarrow}$.
Once the  ODMR resonance is found by scanning the microwave frequency [Fig.~3(b)], we drive the qubit transition on resonance and observe Rabi oscillations [Fig.~3(c)].
Finally, we use Ramsey interference to measure the spin dephasing time $T_2^*$ for both samples [Fig. 3(d)]. 
For \PurifiedSample, which contains a low density of nuclear spins (blue circles), we measure a dephasing time in the range of $T_2^* = 1.5\,\mu$s to $13\,\mu$s. 
For this sample, we find that  $T_2^*$ scales inversely with the qubit frequency $\textrm{f}_{\uparrow\downarrow}$ as shown in Fig.~3(e). 
The observed scaling $T_2^*\propto 1/\textrm{f}_{\uparrow\downarrow}$ indicates that fluctuations of the electronic g-factor $\Delta g$ likely limit the $T_2^*$ via the relation $1/(T_2^*)\propto \Delta g \mu_\textrm{B} B$ where $\mu_\textrm{B}$ is the Bohr magneton. 
Possible origins for $\Delta g$ are discussed in the Supplemental Material~\cite{SI}. 
For \StandardSample{}, which contains a natural abundance of nuclear spins (red squares), we measure $T_2^* \approx 300$\,ns
independent from the magnetic field magnitude
which is similar to typical values observed with \NVm{} centers. 
These results demonstrate that the dephasing time $T_2^*$ of \SiVm\ centers in \StandardSample{} is primarily limited by the nuclear spin bath in the diamond host with a natural abundance of $^{13}$C~\cite{Childress2006}.

Dephasing due to slowly evolving fluctuations in the environment (e.g. nuclear spins) can be suppressed by using dynamical decoupling techniques~\cite{Ryan2010a,DeLange2010a}.
We extend the spin coherence time $T_2$ by implementing Carr-Purcell-Meiboom-Gill (CPMG) sequences with $N=1,2,4,8,16$, and 32 rephasing pulses~\cite{Meiboom1958} in \PurifiedSample{} [Fig.~4(b)]. 
Figure~4(c) shows that the coherence time  increases approximately linearly with the number of rephasing pulses $N$.
The longest observed coherence time is $T_2=13\pm 1.7$\,ms for $N=32$. 
We also implement CPMG sequences for $N=1,2$, and 4 in \StandardSample{} and find similar coherence times $T_2$ as for \PurifiedSample{} [Fig.~4(c)].
We repeat the CPMG measurements at higher temperatures:
at $400$\,mK, the $T_2$ time measured by CPMG2 is identical to $T_2$ at 100\,mK (red and orange data). 
At a temperature of 600 mK, the spin-echo (CPMG1)  $T_2$ is dramatically reduced to 60 $\mu$s.
Spin-echo measurements with \StandardSample{} at a weak magnetic field of 0.2\,kG show high-visibility oscillations  
of the electronic spin coherence~\cite{SI}. These dynamics are suppressed at stronger fields~\cite{SI} and are characteristic of coherent coupling to nearby $^{13}$C nuclear spins~\cite{Childress2006}.

Surprisingly,  the observation in Fig.~4 that the coherence time $T_2$ in both samples is identical for a given $N$ indicates that the coherence time $T_2$ is not limited by the nuclear spin bath, but by another noise source.
This observation is also supported by the approximately linear scaling ($T_2\sim N$)
of coherence with the number of $\pi$-pulses which deviates substantially from the expected $\sim N^{2/3}$ scaling for dipolar coupling to nuclear spins~\cite{Bar-Gill2013,Medford2012}. 
We also do not find a significant difference between $T_2$ measured at different magnetic fields ~\cite{SI},
suggesting that g-factor fluctuations are also not the limiting factor for these measurements.

\begin{figure}[t]
	\includegraphics[width=\linewidth]{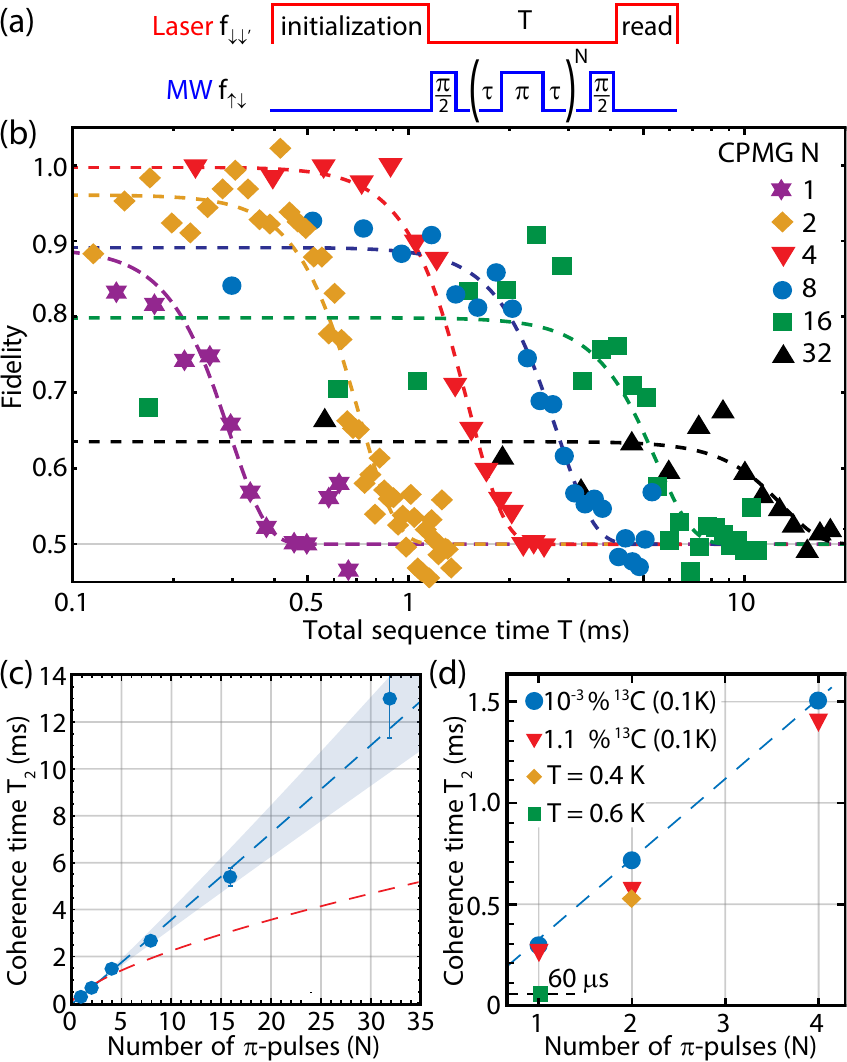}
		\caption{\label{fig4}
	13\,ms spin coherence with dynamical decoupling. 
	(a) CPMG sequence with $N$ rephasing MW $\pi$-pulses. 
	(b) Spin coherence for CPMG sequences with $N$ = 1, 2, 4, 8, 16, and 32 pulses in  \PurifiedSample{} in an aligned magnetic field $B\approx 1.6$\,kG at 100\,mK. 
	The longest measured $T_2$ time is 13\,ms for $N=32$. 
	State fidelity reduces with higher N due to $\pi$-pulse errors.  
	Durations of the initialization and readout laser pulses are $\sim 100$\,ms and $\sim 15$\,ms, correspondingly.
	Dashed lines are fits to $\text{exp}[-(\nicefrac{T}{T_2})^4]$~\cite{SI}.
	(c) $T_2$ coherence vs. number of rephasing pulses $N$ for \PurifiedSample.
	Fitting to $T_2 \propto N^\beta$ gives $\beta = 1.02 \pm 0.05$ (blue dashed line), 
	the shaded region represents a standard deviation of 0.05. 
	For comparison, the red dashed line shows $N^{2/3}$ scaling. 
	(d) $T_2$ coherence vs. number of rephasing pulses $N$ for \PurifiedSample{} and \StandardSample{}. Green and orange points are measured with \StandardSample{} at elevated temperatures.
}
\end{figure}

While the origin of the noise source is at present not understood, the linear dependence of $T_2$ on $N$ suggests that  $T_2$ can potentially be further improved by using additional rephasing pulses. 
In the current measurements, errors due to imperfect $\pi$-pulses ~\cite{SI} result in reduced state fidelities for pulse sequences with $N \geq 32$. 
Pulse errors can be reduced by using decoupling sequences with two-axis ($XY$) control~\cite{DeLange2010a}.
The gate fidelities can also be improved using higher MW Rabi frequencies~\cite{SI} that can be obtained with low-loss superconducting coplanar waveguides~\cite{sigillito2014fast}.

These observations establish the \SiVm\ center as a promising solid-state quantum emitter
for the realization of quantum network nodes using integrated diamond nanophotonics~\cite{Sipahigil2016}. 
Although understanding the noise bath and its effects on the \SiVm{} spin dynamics is an important area of future study, the demonstrated coherence time of 13\,ms is already sufficient to maintain quantum states between 
quantum repeater nodes separated by $10^3$\,km~\cite{Childress2006a}. 
The quantum memory lifetime could be further extended by implementing robust dynamical decoupling 
schemes~\cite{DeLange2010a} or using coherently coupled nuclear spins as longer-lived memories~\cite{Maurer2012a}.
The \SiVm\ spin could also be strongly coupled to localized acoustic~\cite{sohn2017engineering,Burek2016crystal} 
modes by exploiting the large strain  susceptibility of the \SiVm\ centers (PHz / strain)~\cite{sohn2017engineering}. 
This offers
new opportunities for realizing two-qubit gates~\cite{Majer2007,bennett2013phonon, stannigel2011optomechanical} 
and interfacing superconducting quantum circuits with long-lived spin memories and optical photons~\cite{gustafsson2014propagating}.

We thank J. Doyle and D. Patterson for providing cryogenic equipment and expertise, M. Markham, A. Edmonds and D. Twitchen from Element Six Inc. for  providing the samples used in this work,  A. Faraon, J. Kindem and T. Zhong for assistance in the early stages of the experiment and M. Lon\v{c}ar and H. Park for useful discussions. This work was supported by
the DURIP award N00014-15-1-28461234 through ARO, NSF,  CUA,  AFOSR MURI and  ARL. 
Coplanar waveguides were fabricated at the Harvard CNS supported under NSF award ECS-0335765.

{\it Note added.}---Recently, we became aware of 
a complementary experiment by  Becker {\it et. al.}~\cite{Becker2017alloptical}, demonstrating 
all-optical coherent manipulation of the \SiVm{} spin qubit at $\sim 20$\,mK with an observed coherence time 
of $\sim 140$\,ns limited by other impurities in the sample.


%

\end{document}